\documentclass[prd,amsmath,amssymb,showpacs,superscriptaddress,nofootinbib,twocolumn]{revtex4-1}
\usepackage{graphicx}
\usepackage{dcolumn}
\usepackage{bm}
\usepackage{rotating}

\usepackage{overpic}

\usepackage{lineno}
\usepackage{color}

\begin{document}
\normalsize
\parskip=5pt plus 1pt minus 1pt

\title{\boldmath Observation of $J/\psi \rightarrow p\bar{p}a_{0}(980)$ at BESIII}

\author{
M.~Ablikim$^{1}$, M.~N.~Achasov$^{8,a}$, X.~C.~Ai$^{1}$, O.~Albayrak$^{4}$, M.~Albrecht$^{3}$, D.~J.~Ambrose$^{42}$, F.~F.~An$^{1}$, Q.~An$^{43}$, J.~Z.~Bai$^{1}$, R.~Baldini Ferroli$^{19A}$
% ...
\\(BESIII Collaboration)\\
$^{1}$ Institute of High Energy Physics, Beijing 100049, People's Republic of China\\
% ...
$^{3}$ Bochum Ruhr-University, D-44780 Bochum, Germany\\
$^{4}$ Carnegie Mellon University, Pittsburgh, Pennsylvania 15213, USA\\
% ...
$^{8}$ G.I. Budker Institute of Nuclear Physics SB RAS (BINP), Novosibirsk 630090, Russia\\
% ...
$^{19}$ (A)INFN Laboratori Nazionali di Frascati, I-00044, Frascati, Italy; (B)INFN and University of Perugia, I-06100, Perugia, Italy\\
% ...
$^{42}$ University of Rochester, Rochester, New York 14627, USA\\
$^{43}$ University of Science and Technology of China, Hefei 230026, People's Republic of China\\
% ...
$^{a}$ Also at the Novosibirsk State University, Novosibirsk, 630090, Russia\\
% ...
}
\date{\today}
\begin{abstract}
Using $2.25\times10^{8}$ $J/\psi$ events collected with the BESIII
detector at the BEPCII storage rings,
we observe for the first time
the process
$J/\psi\rightarrow p\bar{p}a_{0}(980)$, $a_{0}(980)\rightarrow \pi^{0}\eta$
with a significance of
$6.5\sigma$ ($3.2\sigma$ including systematic uncertainties).
The product branching fraction of
$J/\psi\rightarrow p\bar{p}a_{0}(980)\rightarrow p\bar{p}\pi^{0}\eta$
is measured to be
$(6.8\pm1.2\pm1.3)\times 10^{-5}$, where the first error is statistical and the second is systematic.
This measurement provides information on the $a_{0}$ production near threshold coupling to $p\bar{p}$
and improves the understanding of the dynamics of $J/\psi$ decays to four body processes.
\end{abstract}

\maketitle

\section{\boldmath INTRODUCTION}
As one of the low-lying scalars, the state $a_{0}(980)$ has turned out to be mysterious
in the quark model scenario. Its production near threshold allows tests of various
hypotheses for its structure, including quark-antiquark~\cite{Achasov}, four quarks~\cite{qqqq},
$K\bar{K}$ molecule~\cite{KK} and hybrid states~\cite{qqg}. The measurement of
$J/\psi\rightarrow p\bar{p}a_{0}(980)$ is an additional observable
constraining
any phenomenological models trying to understand the nature of the $a_{0}(980)$.

A chiral unitary coupled channels approach of the Chiral perturbation theory (ChPT)~\cite{Weinberg,Bernard,Pich}
is applied in investigation of the four-body decays $J/\psi\rightarrow N\bar{N}MM$ process~\cite{Chiang}
where the $N$ stands for a baryon and the $M$ for a meson.
In this approach, the
process $J/\psi\rightarrow p\bar{p}\pi^{0}\eta$
is investigated with the $a_{0}(980)$ meson generated through final state interaction (FSI). The amplitude
of this process is calculable except for some coefficients which are not restricted, and its branching fraction
varies within a wide range for different coefficients. Therefore,
an experimental measurement of the process
$J/\psi\rightarrow p\bar{p}a_{0}(980)\rightarrow p\bar{p}\pi^{0}\eta$ is needed for further progress
in understanding of the dynamics of the four-body decay processes
taking the FSI of mesons into account.

In this paper, we present a measurement of $J/\psi\rightarrow p\bar{p}a_{0}(980)$ with $a_{0}(980)$
decaying to $\pi^{0}\eta$ based on $2.25\times10^{8}$ $J/\psi$
events~\cite{Yang} collected with the BESIII detector at BEPCII.

\section{\boldmath THE EXPERIMENT AND DATA SETS}

BESIII/BEPCII \cite{BESIII} is a major upgrade of BESII/BEPC
\cite{Bai}.  BEPCII is a double-ring $e^{+}e^{-}$ collider running at
2.0-4.6 GeV center-of-mass energies; it provides a peak luminosity of
0.4$\times$10$^{33}$~cm$^{-2}$s$^{-1}$ at the center-of-mass energy of
3.097~GeV.

The cylindrical BESIII detector has an effective geometrical acceptance of $93\%$ of 4$\pi$. It contains
a small cell helium-based (40$\%$ He, 60$\%$ C$_{3}$H$_{8}$) main drift chamber (MDC) which has
43 cylindrical layers and provides an average single-hit resolution of 135~$\mu$m and momentum
measurements of charged particles; a time-of-flight system (TOF)
consisting of 5~cm thick plastic
scintillators, with 176 detectors of length 2.4~m in two layers in the barrel and 96 fan-shaped
detectors in the end caps; an electromagnetic calorimeter (EMC) consisting of 6240 CsI(Tl) crystals
in a cylindrical structure and two end caps, which is used to measure the energies of photons and
electrons; and a muon system (MUC) consisting of Resistive Plate Chambers (RPC).
The momentum resolution of the charged particle is 0.5$\%$ at 1~GeV/$c$ in a 1 Tesla magnetic field.
The energy loss ($dE/dx$) measurement provided by the MDC has a resolution of 6$\%$. The time resolution
of the TOF is 80~ps in the barrel detector and 110~ps in the end cap detectors. The energy resolution
of EMC is $2.5\%$ ($5.0\%$) in the barrel (end caps).

Monte Carlo~(MC) simulated events are used to determine the detection efficiency, optimize
selection criteria, and estimate possible backgrounds. The {\sc Geant4}-based~\cite{geant4}
simulation software {\sc Boost}~\cite{Deng} includes the geometric and material description of
the BESIII detectors, the detector response and digitization models, as well as the tracking of
the detector running conditions and performance.
The $J/\psi$ resonance is generated by {\sc kkmc}~\cite{Jadach} which is the event generator
based on precise predictions of the Electroweak Standard Model for the process
$e^{+}e^{-}\rightarrow f\bar{f}+n\gamma$, where $f=e,~\mu,~\tau,~u,~d,~c,~s,~b$ and $n$
is an integer number $\ge 0$. The subsequent decays are generated with {\sc EvtGen}~\cite{Ping}
with branching fractions being set to the world average values according to the Particle
Data Group~(PDG) \cite{PDG} and the remaining unmeasured decays are generated by
{\sc Lundcharm}~\cite{lund}.
A sample of $2.25\times10^8$ simulated events, corresponding to the
luminosity of data, is used to study background processes from
$J/\psi$ decays (`inclusive backgrounds').
A signal MC sample with more than
10 times of the observed events in data for the process $J/\psi\rightarrow
p\bar{p}a_{0}(980)\rightarrow p\bar{p}\pi^{0}\eta$  is generated,
where the shape of the $a_{0}(980)$ is parameterized with the Flatt\'e
formula~\cite{Flatte}.

\section{\boldmath  EVENT SELECTION}
We select the process $J/\psi\rightarrow p\bar{p}\pi^{0}\eta$, with
both $\pi^{0}$ and $\eta$ decaying to two photons, for this analysis.
A good charged track is required to have
good quality in the track fitting and be within the polar angle coverage of the MDC,
i.e., $|\cos\theta|~<0.93$, and pass within 1~cm of the $e^{+}e^{-}$ interaction point
in the transverse direction to the beam line and within 10~cm of the interaction point
along the beam axis. Since the charged track in this process has relatively low transverse
momentum, charged particle identification~(PID) is only based on the $dE/dx$ information
with the confidence level Prob$_{\textrm{PID}}$(i) calculated for each particle hypothesis
$i$ ($i=\pi/K/p$). A charged track with Prob$_{\textrm{PID}}$(p)$>$Prob$_{\textrm{PID}}(K)$
and Prob$_{\textrm{PID}}$(p)$>$Prob$_{\textrm{PID}}(\pi)$ is identified as a proton or an
antiproton candidate. Photon candidates are required to have a minimum energy deposition
of 25~MeV in the barrel ($|\cos\theta|<$0.8) of the EMC and 50~MeV in the end caps~(0.86$<|\cos\theta|<$0.92)
of the EMC. EMC timing requirements ($0\leq T\leq14$ in units of 50~ns) are used to
suppress electronic noise and to remove showers unrelated to
the event.
At the event selection
level, candidate events are required to have at least two good charged tracks with one proton
and one antiproton being identified, and at least four good photons.

\begin{figure*}[htbp]
\centering
\begin{overpic}[width=7.4cm,height=6.3cm,angle=0]{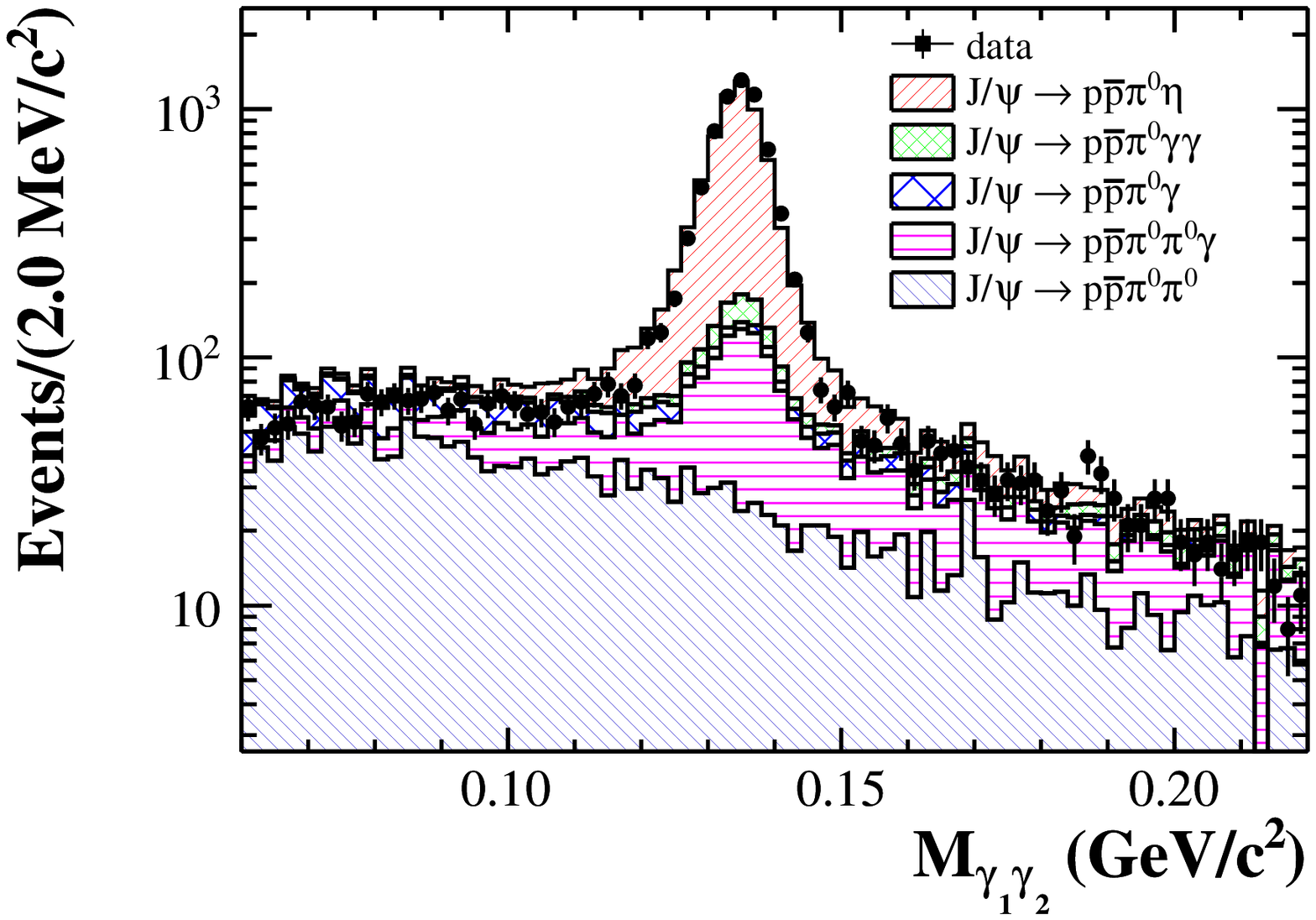}
\put(25,65){\large\bf (a)}
\end{overpic}
\begin{overpic}[width=7.4cm,height=6.3cm,angle=0]{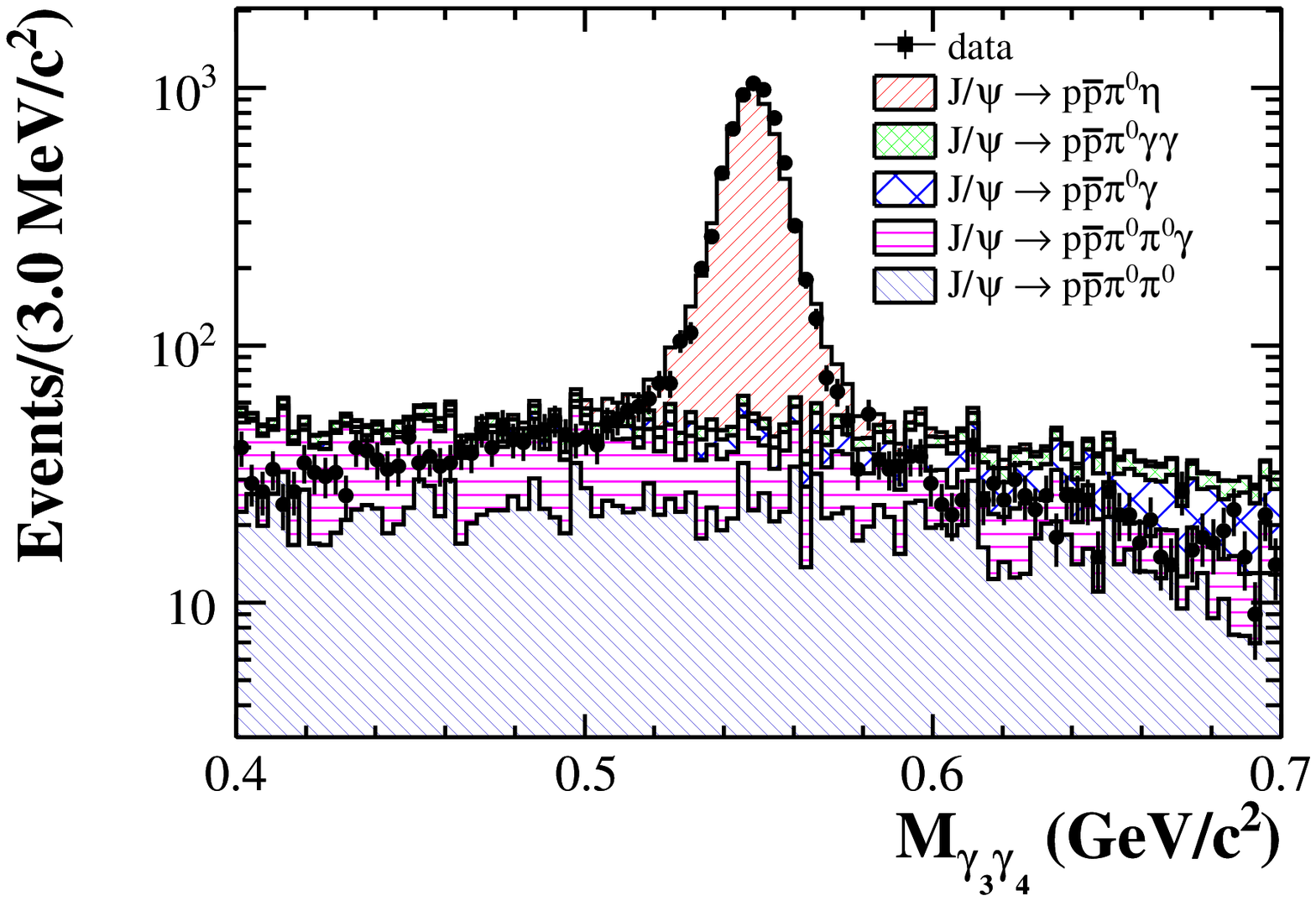}
\put(25,65){\large\bf (b)}
\end{overpic}
\vskip -0.3cm
\parbox[1cm]{16cm} {
\caption{ The invariant mass distribution of (a) $\pi^{0}$ candidates and
   (b) $\eta$ candidates. Dots with error bars are data. The hatched histograms
   are processes with different final states from simulated $J/\psi$ decays.}
 \label{gamgam1}
}
\end{figure*}

We then perform a kinematic fit which imposes energy and momentum conservation at the
production vertex to
combinations of one
proton and
one
antiproton candidate and four photons. For events with more
than four photons, we consider all possible four-photon combinations, and the one giving the
smallest $\chi^{2}_{4C}$ for the kinematic fit is selected for further analysis. To improve
the
signal-to-background ratio, events with $\chi^{2}_{4C}<$35 are accepted; this
optimizes the figure of merit $S/\sqrt{S+B}$, where $S$ and $B$ are the numbers of MC simulated signal
and inclusive background events respectively.
The best photons pairing to $\pi^{0}$ and $\eta$ in the four selected photons are selected by
choosing the combination that gives the minimum $\chi^{2}$-like variable
$$\chi^{2}_{\pi^{0}\eta}=\frac{(M_{\gamma_{1}\gamma_{2}}-M_{\pi^{0}})^{2}}{\sigma^{2}_{{\pi^{0}}}}
+\frac{(M_{\gamma_{3}\gamma_{4}}-M_{\eta})^{2}}{\sigma^{2}_{{\eta}}},$$
where $M_{\gamma\gamma}$ is the invariant mass of two photons after
kinematic fit and $M_{\pi^{0}/\eta}$ is the $\pi^{0}/\eta$ mass from PDG~\cite{PDG}.
The mass resolutions for the $\pi^{0}$ and $\eta$, $\sigma_{\pi^{0}}$ and
$\sigma_{\eta}$ are extracted by fitting the corresponding mass
spectra in the signal MC sample;  they are found to be 6.0~MeV/$c^{2}$ and
9.8~MeV/$c^{2}$ respectively.
A MC study shows the rate of correct combination of photons is greater than $99\%$ by using the $\chi^{2}_{\pi^{0}\eta}$ metric.
To suppress $p\bar{p}\pi^{0}\pi^{0}$ final states surviving in the 4C fit, we select two-photon
pairs giving a minimum $\chi^{2}_{\pi^{0}\pi^{0}}=\frac{(M_{{\gamma_{1}\gamma}_{2}}-M_{\pi^{0}})^{2}}
{\sigma^{2}_{\pi^{0}}}+\frac{(M_{{\gamma_{3}\gamma_{4}}}-M_{\pi^{0}})^{2}}{\sigma^{2}_{\pi^{0}}}$
and reject events with $\chi^{2}_{\pi^{0}\pi^{0}}$ less than 100. Figure~\ref{gamgam1} shows the
mass spectra of selected $\gamma\gamma$ pairs for data and MC, where $\gamma_{1}\gamma_{2}$
indicates $\pi^{0}$ candidates and $\gamma_{3}\gamma_{4}$ indicates $\eta$ candidates. The
hatched histograms represent MC shapes from backgrounds and signal, where the background shapes
are normalized based on their branching fractions and the signal shape is normalized to the
rest area of the histogram of the data. We then require the mass of $\pi^{0}$ and $\eta$
candidates
to be within a 3$\sigma$ window
around
their mean values.

\section{\boldmath DATA ANALYSIS}

\begin{figure*}[htbp]
\centering
\begin{overpic}[width=7.4cm,height=6.3cm,angle=0]{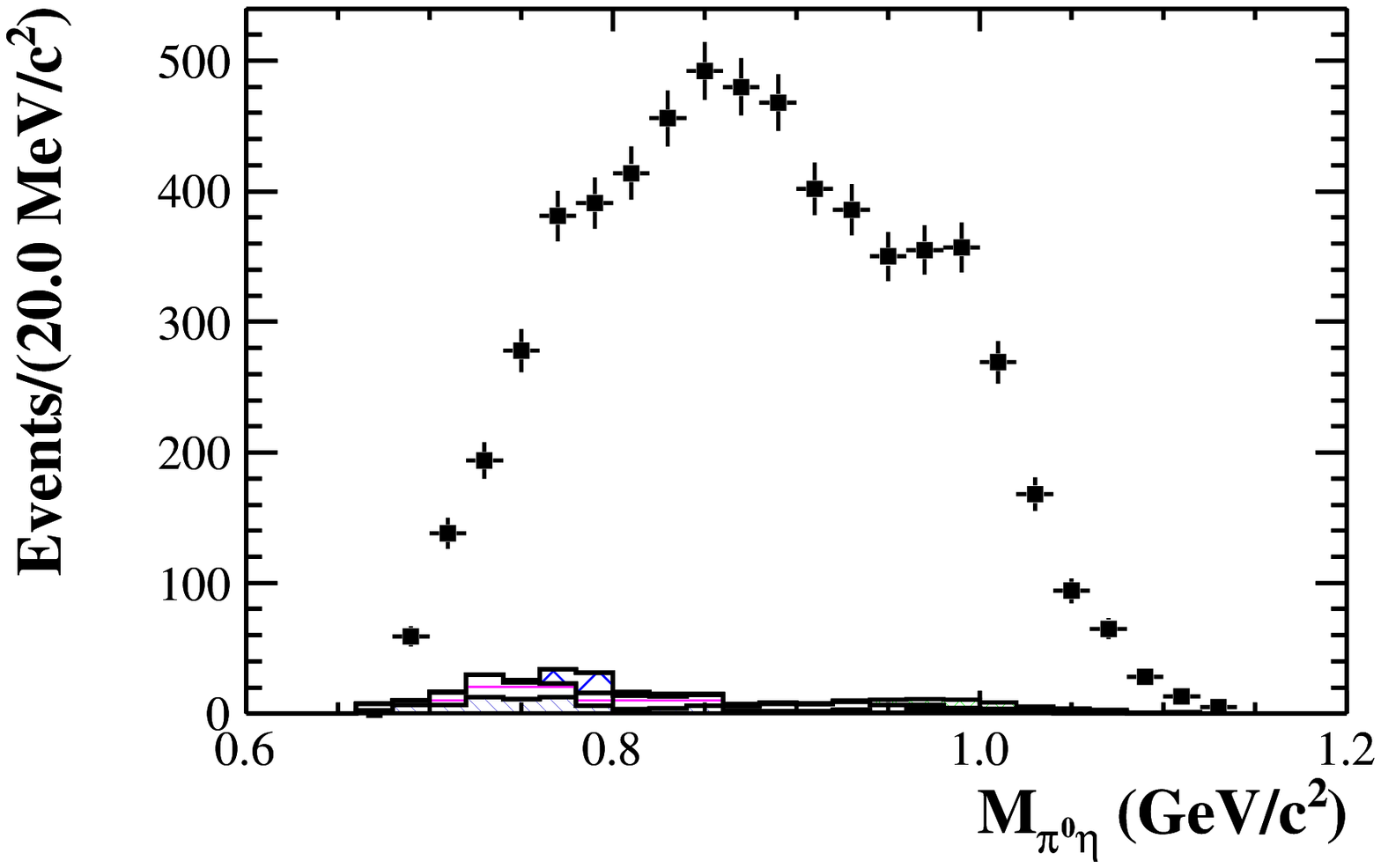}
\put(25,65){\large\bf (a)}
\end{overpic}
\begin{overpic}[width=7.4cm,height=6.3cm,angle=0]{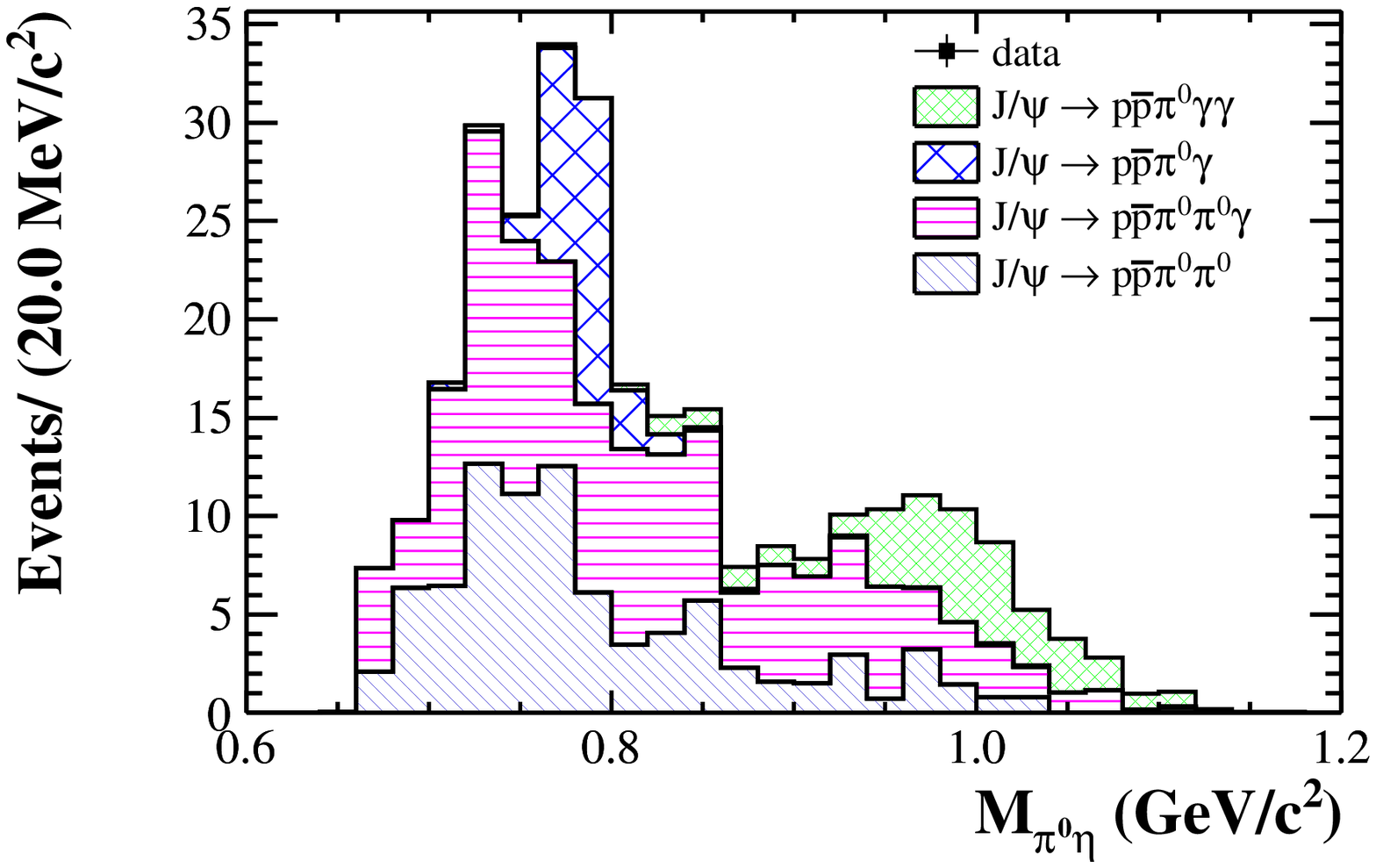}
\put(25,65){\large\bf (b)}
\end{overpic}
\vskip -0.3cm
\parbox[1cm]{16cm} {
\caption{ (a) The mass spectrum of $\pi^{0}\eta$ for data and exclusive backgrounds.
The dots with error bars represent data and the others are exclusive backgrounds after normalization.
(b) The mass spectra of $\pi^{0}\eta$ for exclusive backgrounds.}
 \label{pi0eta}
}
\end{figure*}

The backgrounds contaminating the
selected $J/\psi\rightarrow p\bar{p}\pi^{0}\eta$ candidates arise
mainly from events with the same topology ($p\bar{p}\gamma\gamma\gamma\gamma$), events with
an additional undetected photon ($p\bar{p}\gamma\gamma\gamma\gamma\gamma$), and events with a fake photon
being reconstructed ($p\bar{p}\gamma\gamma\gamma$). The potential final states of background are
categorized into four kinds: $p\bar{p}\pi^{0}\pi^{0}$, $ p\bar{p}\pi^{0}\pi^{0}\gamma$,
$p\bar{p}\pi^{0}\gamma $ and $ p\bar{p}\pi^{0}\gamma\gamma$, where the $p\pi^{0}$ can be produced
from intermediate states $\Sigma$ or $\Delta$, and $\gamma \pi^{0}$
can be produced from $\omega$.
Since the branching fractions for the exclusive background processes
$J/\psi\rightarrow\Sigma^{+}\Sigma^{-}(\gamma)/
\Delta^{+}\Delta^{-}(\gamma)/p\bar{p}\omega(n\gamma)$
have not yet been measured, we determine them from the same $J/\psi$
data sample.
The measurements are performed by requiring different numbers of
photon candidates in one event and selecting the
combination of $p\pi^{0}$ with invariant mass closest to the mass of $\Sigma$ or $\Delta$, or selecting
the combination of $\gamma\pi^{0}$ closest to the mass of
$\omega$. The
measured
branching fractions are shown in Table~\ref{bkg}, where the uncertainty is statistical only.
With the detection efficiency correction for the exclusive background satisfying the $p\bar{p}\pi^{0}\eta$
selection criteria, the contribution of the exclusive backgrounds is calculated to be $290\pm19$,
which accounts for $4.3\%$ of the surviving
events found in
data. The distributions of $M_{\pi^{0}\eta}$ for data
and backgrounds after normalization are presented in Fig.~\ref{pi0eta}. A structure around
1.0 GeV (Fig.~\ref{pi0eta}(a)) in data is
clearly visible,
but
is not seen
significantly in the
corresponding distribution of the exclusive backgrounds (Fig.~\ref{pi0eta}(b)).

\begin{table*}[htbp]
\caption{Backgrounds of the final states with $p\bar{p}\pi^{0}\pi^{0}$, $p\bar{p}\pi^{0}\pi^{0}\gamma$,
$p\bar{p}\pi^{0}\gamma$ and $p\bar{p}\pi^{0}\gamma\gamma$, where $Br$ is the branching fraction of
each channel, with statistical error only, $\varepsilon_{MC}^{sel}$ is the selected efficiency of
each channel determined with 50k MC sample,  and $N^{Norm}$ is the number of background events
normalized to the total $J/\psi$ data.
}
\begin{center}
\begin{tabular}{cccc}
\hline
\hline
Channel($J/\psi\rightarrow$) & Br & $\varepsilon_{MC}^{sel}$ & $N^{Norm}$ \\
\hline
$p\bar{p}\pi^{0}\pi^{0}$ & $(1.60\pm0.26)\times10^{-3}$ & $1.68\times10^{-4}$ & $61\pm10$ \\
$\Sigma^{+}\Sigma^{-}\rightarrow p\pi^{0}\bar{p}\pi^{0}$  & $(2.77\pm0.03)\times10^{-4}$ & $1.26\times10^{-4}$ & $8\pm0$\\
$\Delta^{+}\Delta^{-}\rightarrow p\pi^{0}\bar{p}\pi^{0}$ & $(2.30\pm0.07)\times10^{-4}$ & $1.76\times10^{-4}$ & $9\pm0$ \\
$p\pi^{0}\Delta^{-}+c.c\rightarrow p\pi^{0}\bar{p}\pi^{0}$ & $(2.04\pm0.06)\times10^{-4}$ & $1.76\times10^{-4}$ & $8\pm0$\\
$\gamma\Sigma^{+}\Sigma^{-}\rightarrow \gamma p\pi^{0}\bar{p}\pi^{0}$  & $(3.31\pm0.12)\times10^{-5}$ & $2.98\times10^{-3}$ & $23\pm1$ \\
$\gamma\Delta^{+}\Delta^{-}\rightarrow \gamma p\pi^{0}\bar{p}\pi^{0}$ & $(5.40\pm0.50)\times10^{-5}$ & $2.86\times10^{-3}$ & $35\pm3$ \\
$\gamma p\pi^{0}\Delta^{-}+c.c\rightarrow \gamma p\pi^{0}\bar{p}\pi^{0}$ & $(14.40\pm2.80)\times10^{-5}$ & $2.44\times10^{-3}$ & $78\pm15$ \\
$p\bar{p}\omega\rightarrow p\bar{p}\gamma\pi^{0}$  & $(9.11\pm1.27)\times10^{-5}$ & $1.59\times10^{-3}$ & $33\pm5$ \\
$\gamma p\bar{p}\omega\rightarrow \gamma p\bar{p}\gamma\pi^{0}$ & $(1.28\pm0.07)\times10^{-5}$ & $1.14\times10^{-2}$ & $33\pm2$ \\
\small $J/\psi\rightarrow p\bar{p}\eta',\eta'\rightarrow \gamma\omega, \omega\rightarrow\gamma\pi^{0}$ & \small $(4.78\pm0.99)\times10^{-7}$ & \small $1.80\times10^{-2}$  & $2\pm0$\\
Total  &  &   & $290\pm19$ \\
\hline
\hline
\end{tabular}
\label{bkg}
\end{center}
\end{table*}

\begin{figure*}[htbp]
\centering
\begin{overpic}[width=7.4cm,height=6.3cm,angle=0]{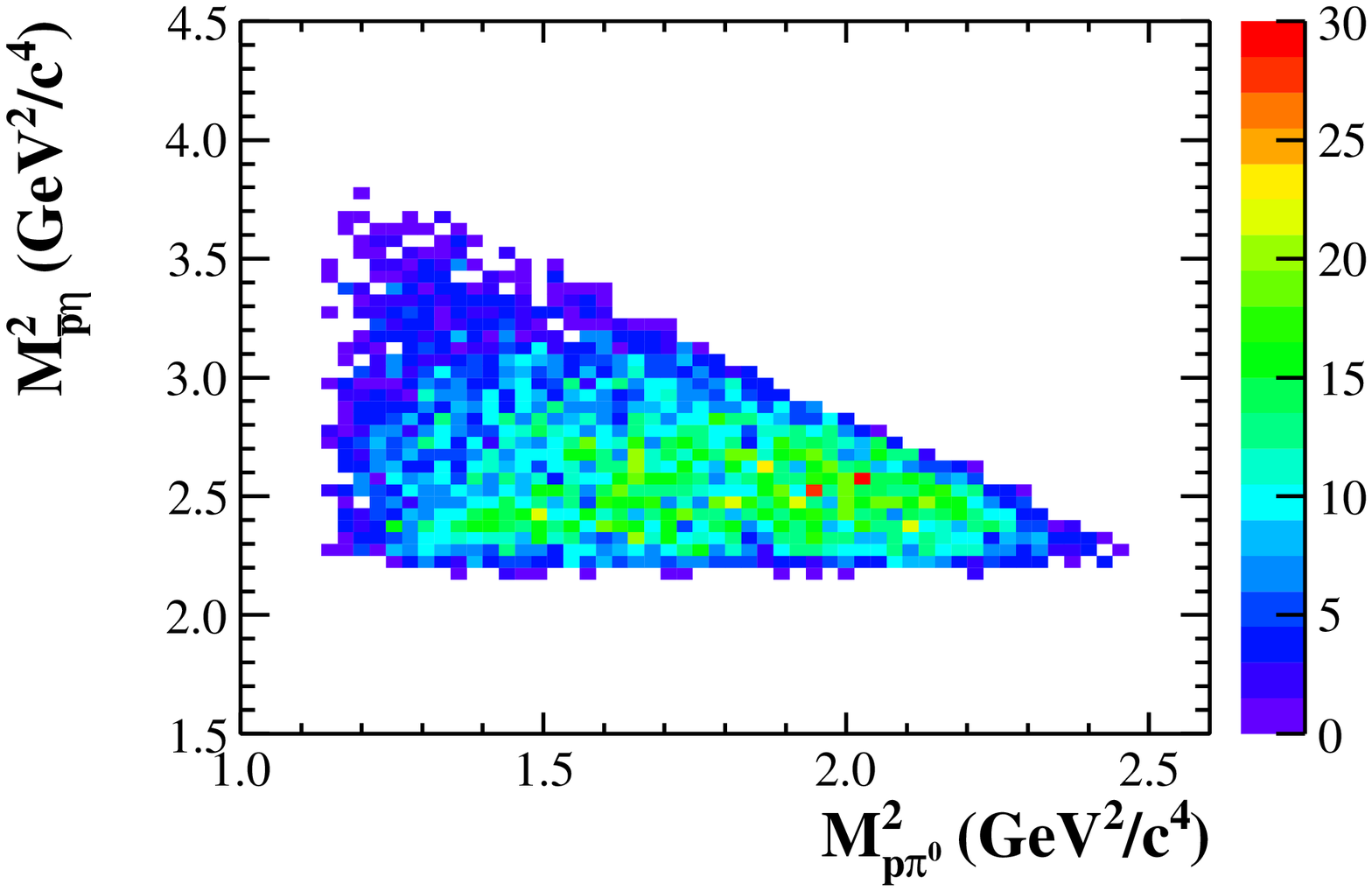}
\put(25,65){\large\bf (a)}
\end{overpic}
\begin{overpic}[width=7.4cm,height=6.3cm,angle=0]{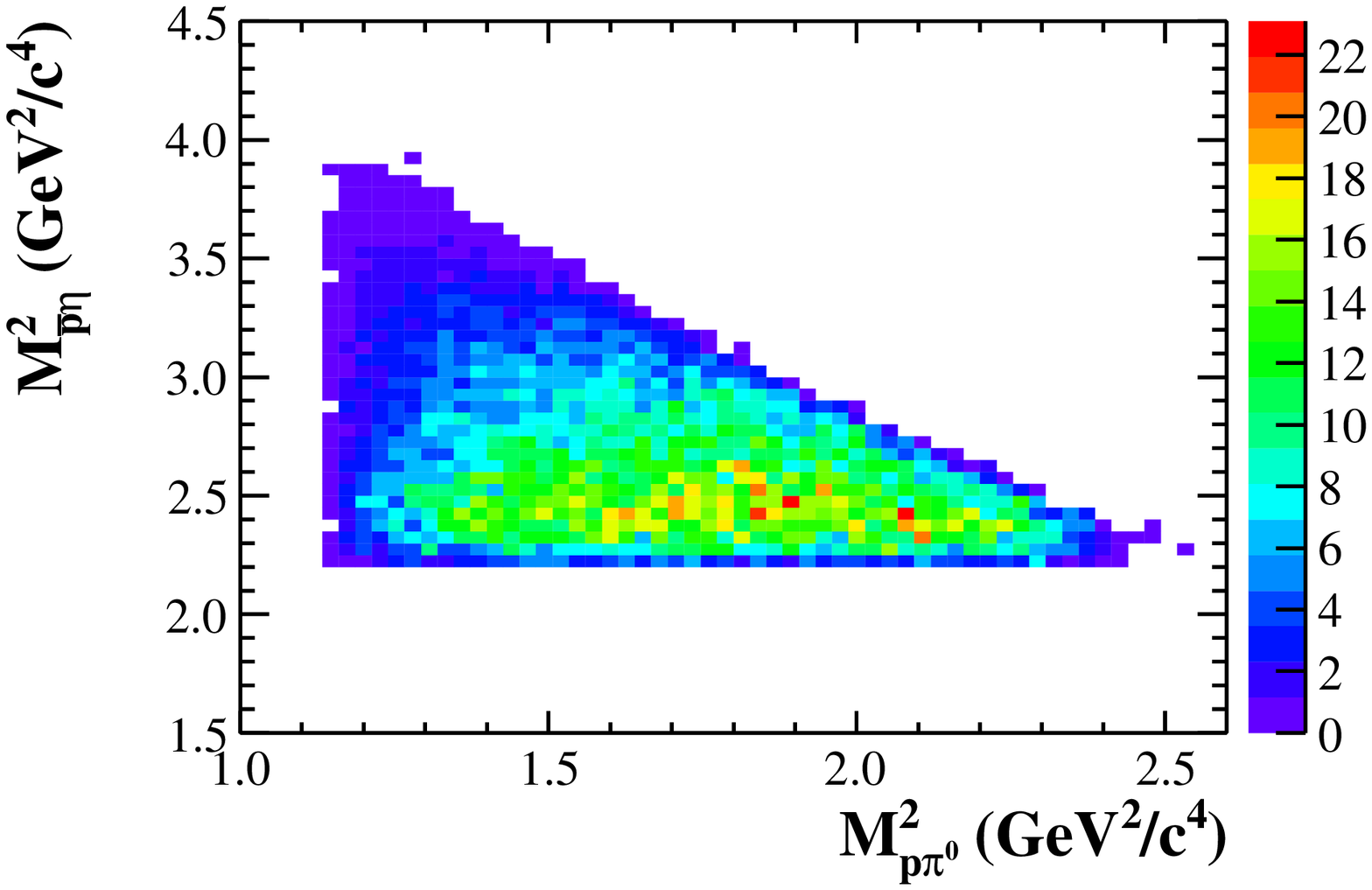}
\put(25,65){\large\bf (b)}
\end{overpic}
\vskip -0.3cm
\parbox[1cm]{16cm} {
\caption{ (a) The scatter plot of $M^{2}_{p\pi^{0}}$ versus $M^{2}_{\bar{p}\eta}$
from data. (b) The scatter plot of $M^{2}_{p\pi^{0}}$ versus $M^{2}_{\bar{p}\eta}$
from MC projection of all intermediate states superimposed.}
 \label{scatter}
}
\end{figure*}

The studies of the mass spectra of $M_{p\pi^{0}}$ and $M_{p\eta}$ show that the processes
with intermediate states of $N(1440)$, $N(1535)$ and $N(1650)$ are the dominant contributions
to $J/\psi\rightarrow p\bar{p}\pi^{0}\eta$ where $N(1440)$ decays to $p\pi^{0}$, $N(1535)$
decays to $p\pi^{0}$ or $p\eta$, and $N(1650)$ decays to $p\eta$, with the charge-conjugate
modes being implied. A simple partial wave analysis (PWA) by calculating the amplitudes of
these processes according to their Feynman Diagrams~\cite{FDC} is applied to the surviving
events in data. The maximum likelihood method is used to fit the branching fraction of these
intermediate states and their interferences. Figure~\ref{scatter}(a) shows the scatter plot
of $M^{2}_{p\pi^{0}}$ versus $M^{2}_{\bar{p}\eta}$ in data, which is consistent with the
scatter plot of $M^{2}_{p\pi^{0}}$ versus $M^{2}_{\bar{p}\eta}$ of the best fit result
shown in Fig.~\ref{scatter}(b). The interference between the
processes with $N^{*}$ and the $p\bar{p}a_{0}(980)$ is found to be very small and is neglected
in the following. The yield of $J/\psi\rightarrow p\bar{p}a_{0}(980)\rightarrow p\bar{p}\pi^{0}\eta$
obtained by the
PWA is within $1\sigma$ statistical deviation of that
obtained
by fitting the mass spectrum of $\pi^{0}\eta$
described below.
When
applying the PWA without the component $J/\psi\rightarrow p\bar{p}a_{0}(980)$,
no enhancement around 1.0 GeV is observed in the MC projection of $\pi^{0}\eta$ mass spectrum, which
indicates that the enhancement seen in data is not from the processes with $N^{*}$ intermediate states
or their interferences.

An unbinned
extended
maximum likelihood fit is performed on the $\pi^{0}\eta$ mass spectrum.
The probability density function (PDF) is
$$F(m) = f_\mathrm{sig}\,\sigma(m)\otimes(\varepsilon(m)\times \hat{T}(m)) + (1-f_\mathrm{sig})\,B(m).$$

Here, $f_\mathrm{sig}$ is the fraction of $p \bar{p} a_0(980)$ signal events.
The signal shape of $a_{0}(980)$ is described as an efficiency-weighted Flatt\'e formula
($\varepsilon(m)\times \hat{T}(m)$)  convoluted with a resolution function $\sigma(m)$.
The
non-$a_0(980)$
background shape, expressed by $B(m)$, is described by a third-order Chebychev
polynomial function. The Flatt\'e formula~\cite{Flatte} is used to parameterize
the $a_{0}(980)$ amplitudes coupling to $\pi^{0}\eta$ and $K\bar{K}$ by a two-channel
resonance expressed as
$$\hat{T}(m)
\propto
  \frac{1}{(m_{a_{0}}^{2}-m^{2})^{2}+
    (\rho_{\pi^{0}\eta}g_{a_{0}\eta\pi^{0}}^{2}+\rho_{K\bar{K}}g_{a_{0}K\bar{K}}^{2})^{2}},$$
where $\rho_{\pi^{0}\eta}$ and $\rho_{K\bar{K}}$ are the decay momenta of the
$\pi^{0}$ or $K$ in the $\pi^{0}\eta$ or $K\bar{K}$ rest frame, respectively.
The two coupling constants
$g_{a_{0}\pi^{0}\eta}$ and $g_{a_{0}K\bar{K}}$  stand for $a_{0}(980)$ resonance coupling
to $\pi^{0}\eta$ and $K\bar{K}$, respectively. The experiment results from Refs.~\cite{BNL,KLOE2,CB2}
are consistent with each other and the weighted average of them are calculated as
$g_{a_{0}\pi^{0}\eta}= 2.83\pm0.05$ and $g_{a_{0}K\bar{K}}=2.11\pm0.06$. In the fit,
the two coupling constants $g_{a_{0}\pi^{0}\eta}$ and $g_{a_{0}K\bar{K}}$ are fixed to
2.83 and 2.11, respectively.

The mass-dependent efficiency $\varepsilon(m)$ is studied by using a large
phase space MC $J/\psi\rightarrow p\bar{p}\pi^{0}\eta$ sample, where the efficiency curve derived
from the four-body phase space MC is compatible with that from signal MC of $p\bar{p}a_{0}(980)$.
The detector resolution $\sigma(m)$
of $M_{\pi^{0}\eta}$ is extracted by using a large sample of simulated signal
events $J/\psi\rightarrow p\bar{p}a_{0}(980), a_{0}(980)\rightarrow \pi^{0}\eta$,
with the width of the $a_{0}(980)$ set to zero.

\begin{figure}[htbp]
\begin{center}
\begin{overpic}[width=8.5cm,height=7.24cm,angle=0]{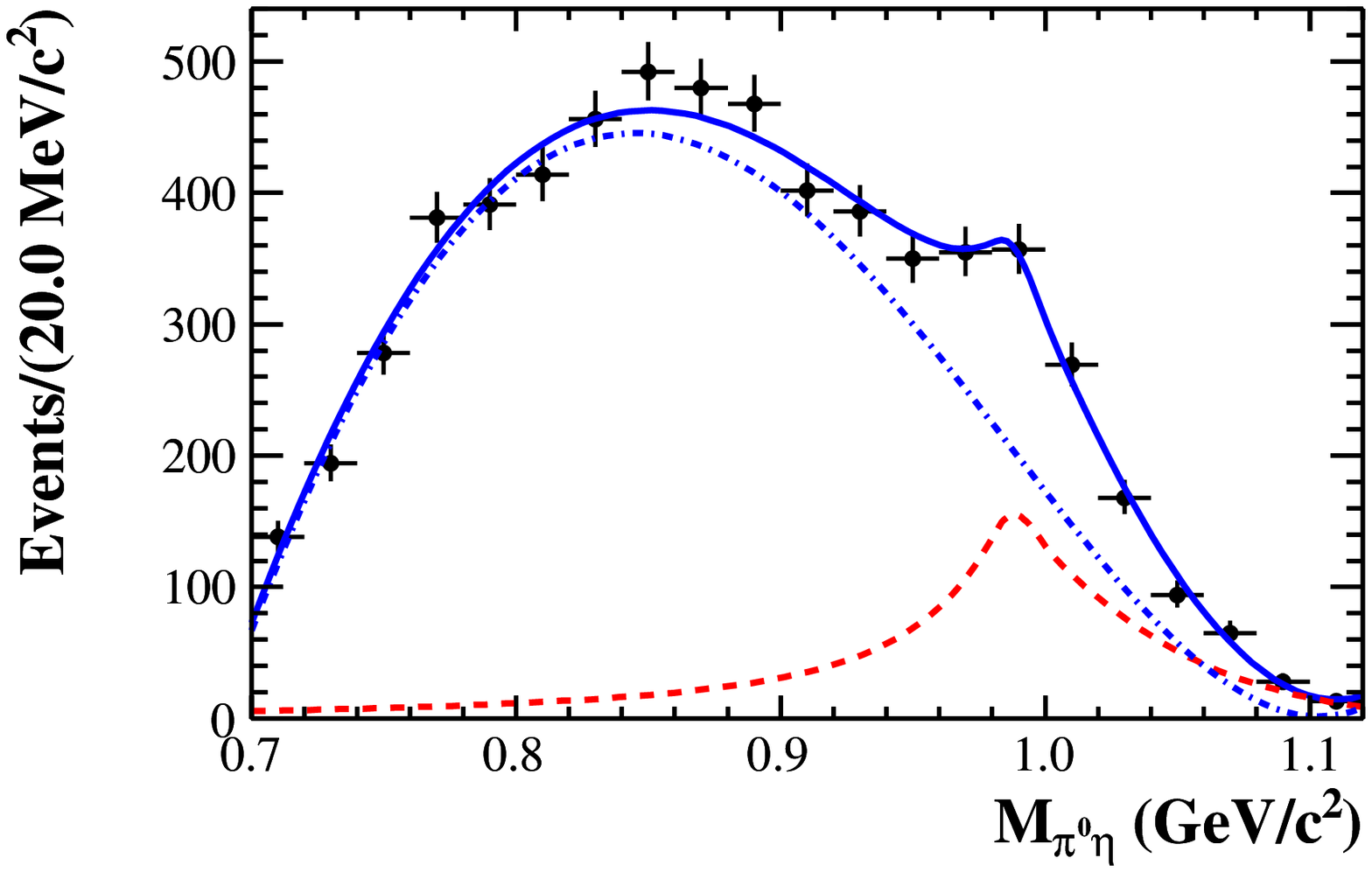}
\end{overpic}
\caption{   The results of fitting the mass spectrum for $\pi^{0}\eta$.
Dots with error bars are data and the solid line is the fitted spectrum.
The dash-dotted line shows
the non-$a_0(980)$
background described by a third-order Cheybechev
polynomial. The dashed line shows the signal described by an efficiency-weighted
Flatt\'e formula convoluted with a resolution function. }\label{fita0}
\end{center}
\end{figure}

In the fit, the signal fraction $f_\mathrm{sig}$, the $a_0(980)$ mass, and the parameters of the
background polynomial are allowed to vary.
The fit result of $M_{\pi^{0}\eta}$ is shown in Fig.~\ref{fita0}. The yield of $a_{0}(980)$
events is $849\pm144$, with a statistical significance of 6.5$\sigma$ which is calculated from
the log-likelihood difference between fits with and without the $a_{0}(980)$ signal component.
The fit mass is $1.012\pm0.007$~GeV/$c^{2}$, which is slightly higher than the PDG value~\cite{PDG}.
The robustness of this result has been validated with a toy MC study. Different signal MC
samples of $J/\psi\rightarrow p\bar{p}a_{0}(980), a_{0}(980)\rightarrow \pi^{0}\eta$ are
generated with different mass and width of the $a_{0}(980)$. Background events are randomly sampled
according to the background shapes.
In all cases, the fit value
\color{black}
of the $a_{0}(980)$ mass is found to be consistent
with the input value within statistical uncertainties. The product branching fraction
$Br(J/\psi\rightarrow p\bar{p}a_{0}(980)\rightarrow p\bar{p}\pi^{0}\eta)$ is calculated to be
$(6.8\pm1.2)\times10^{-5}$, where the uncertainty is statistical only.

\section{\boldmath ESTIMATION OF SYSTEMATIC UNCERTAINTIES}

The systematic uncertainties on the measurement of $Br(J/\psi\rightarrow p\bar{p}a_{0}(980)\rightarrow
p\bar{p}\pi^{0}\eta)$ are summarized in Table~\ref{sys}.

\begin{table}[htbp]
\centering
\caption{Summary of systematic uncertainties on $Br(J/\psi\rightarrow p\bar{p}a_{0}(980)\rightarrow p\bar{p}\pi^{0}\eta)$.}
\begin{tabular}{lc}
\hline
\hline
Source & Uncertainty \\
\hline
Tracking & $9.0\%$  \\
Particle identification & $4.0\%$ \\
Photon detection & $4.0\%$ \\
4C kinematic fitting & $3.2\%$\\
$\chi^{2}_{\pi^{0}\pi^{0}}$ cut & $1.3\%$ \\
Coupling constants &  $3.8\%$  \\
Fit range & $9.2\%$\\
Background shape & $12.6\%$\\
Number of $J/\psi$ events & $1.2\%$\\
Total &  $19.6\%$ \\
\hline
\hline
\end{tabular}
\label{sys}
\end{table}

Systematic uncertainties due to tracking and PID efficiency, photon detection efficiency, the kinematic
fit and the $\pi^0\pi^0$ veto arise due to imperfect modelling of the data by the simulation.
The systematic uncertainty associated with the tracking efficiency as a function
of transverse momentum and the uncertainty due to the PID efficiency of proton/antiproton
have been studied by a control sample of $J/\psi\rightarrow p\bar{p}\pi^{+}\pi^{-}$ decays using a
technique similar to that discussed in Ref.~\cite{photon}. In this paper, due to the
low transverse momentum of proton and antiproton, the uncertainty of tracking efficiency
is determined by the weighted uncertainty $\Sigma_{i}\varepsilon_{i}r_{i}$, where
$\varepsilon_{i}$ represents the data/MC difference in each transverse momentum
bin~\cite{photon} and $r_{i}$ represents the proportion of each transverse momentum
bin in data. The systematic uncertainty due to the tracking efficiency is estimated to be
$4.0\%$ per proton and $5.0\%$ per antiproton, respectively.
 The large uncertainty of tracking efficiency is because of limited statistics in control sample and improper simulation of interactions with material for low momentum proton and antiproton.
 The uncertainty due to PID
efficiency is $2.0\%$ per proton or antiproton.

The systematic uncertainty due to photon detection is $1.0\%$ per photon. This is
determined from studies of the photon detection efficiency in the control sample
$J/\psi\rightarrow \rho^{0}\pi^{0}$~\cite{photon}.

To estimate the uncertainty from the kinematic fit, the efficiency of
the selection on the $\chi^2_{4C}$ of the
kinematic fit
is studied using events of the decay $J/\psi\rightarrow p\bar{p}\eta$, $\eta\rightarrow \pi^{0}\pi^{0}\pi^{0}$.
The uncertainty
associated with
the kinematic fit is determined by the difference of efficiencies for MC and
data, and is estimated to be $3.2\%$ for $\chi^{2}_{4C}<35$.

 The systematic uncertainty arising from the $\pi^{0}\pi^{0}$ veto metric ($\chi^{2}_{\pi^{0}\pi^{0}}>100$)
is studied by a control sample $J/\psi\rightarrow \omega\eta\rightarrow \pi^{+}\pi^{-}\pi^{0}\eta$. The control sample
is selected due to its similar final states to signal, high statistics, and narrow $\omega$/$\eta$ signals to extract the
efficiency precisely.
To better model the signal process $J/\psi\rightarrow p\bar{p}a_{0}(980)\rightarrow p\bar{p}\pi^{0}\eta$, the
$\chi^{2}_{\pi^{0}\pi^{0}}$ distribution of control sample is weighted to that of signal process.
The event number of control sample is extracted by fitting invariant mass of $\pi^{+}\pi^{-}\pi^{0}$ with a double
Gaussian function, and the efficiency for $\chi^{2}_{\pi^{0}\pi^{0}}$ requirement is ratio of the number of events
that with and without veto metric, to be $(97.4\pm1.0)\%$ and $(97.6\pm0.4)\%$ for data and MC, respectively,
where the errors are statistical only.
Conservatively, the systematic uncertainty of $\chi^{2}_{\pi^{0}\pi^{0}}$ veto metric is estimated to be $1.3\%$.

The systematic uncertainty due to the signal shape is determined by varying the
coupling constants by 1$\sigma$ within their center values for $g_{a_{0}\pi^{0}\eta}$
and $g_{a_{0}K\bar{K}}$ separately. The largest difference is taken as the
uncertainty.

To study the uncertainty from background, alternative background shapes are obtained
by varying the fitting range from [0.7, 1.12] GeV/c$^{2}$ to [0.73, 1.12] GeV/c$^{2}$
and changing order of Chebychev polynomial from third-order to fourth-order, which
introduce uncertainties of $9.2\%$ and $12.6\%$, respectively.

The systematic uncertainty of the total number of $J/\psi$ events is obtained
by studying inclusive hadronic $J/\psi$ decays~\cite{Yang} to be $1.2\%$.

We treat all the sources of systematic uncertainties as uncorrelated and sum them
in quadrature to obtain the total systematic uncertainty.

\section{\boldmath CONCLUSION AND DISCUSSION}

Based on $2.25\times10^{8}$  $J/\psi$ events collected with the BESIII detector
at BEPCII,  we observe $J/\psi\rightarrow p\bar{p}a_{0}(980)$, $a_{0}(980)\rightarrow \pi^{0}\eta$
for the first time with a statistical significance of 6.5$\sigma$.
Taking the systematic uncertainty into account, the significance is $3.2\sigma$.
Without considering the
interference between the signal channel and the same final states with intermediate $N^{*}$
states, the branching fraction is measured to be
$$Br(J/\psi\rightarrow p\bar{p}a_{0}(980)\rightarrow p\bar{p}\pi^{0}\eta)
= (6.8\pm1.2\pm1.3)\times10^{-5},$$
where the first uncertainty is statistical and the second is systematic.

Our measurement provides a quantitative comparison with the chiral unitary approach~\cite{Chiang}.
This approximation uses several coefficients in the parametrization of meson-meson amplitudes.
One of them, namely $r_{4}$ in \cite{Chiang}, is constrained by fitting the
$\pi^{+}\pi^{-}$ invariant mass
distribution in the decay $J/\psi\rightarrow p\bar{p}\pi^{+}\pi^{-}$;
the fit suggests two equally
possible values, $r_{4}=0.2$ and $r_{4}=-0.27$. The theory also predicts that the branching
fractions of $J/\psi\rightarrow p\bar{p}a_{0}(980)$ and $J/\psi\rightarrow p\bar{p}\pi^{+}\pi^{-}$
are comparable for $r_{4}=-0.27$,
while the
branching fraction of the former is one or two orders of
magnitude lower than that of the latter for $r_{4}=0.2$. Taking the branching fraction of
$J/\psi\rightarrow p\bar{p}\pi^{+}\pi^{-}$ from PDG~\cite{PDG}, the ratio of
$Br(J/\psi\rightarrow p\bar{p}a_{0}(980)\rightarrow p\bar{p}\pi^{0}\eta)$ to
$Br(J/\psi\rightarrow p\bar{p}\pi^{+}\pi^{-})$ is
found to be
about $10^{-2}$, which shows preference to $r_{4}=0.2$.

\acknowledgments
The BESIII collaboration thanks the staff of BEPCII and the computing
center for their strong support. This work is supported in part by the
Ministry of Science and Technology of China under Contract
No. 2009CB825200; Joint Funds of the National Natural Science
Foundation of China under Contracts Nos. 11079008, 11179007, U1332201;
National Natural Science Foundation of China (NSFC) under Contracts
Nos. 10625524, 10821063, 10825524, 10835001, 10935007, 11125525,
11235011, 11335008, 11275189, 11322544, 11375170; the Chinese Academy
of Sciences (CAS) Large-Scale Scientific Facility Program; CAS under
Contracts Nos. KJCX2-YW-N29, KJCX2-YW-N45; 100 Talents Program of CAS;
German Research Foundation DFG under Contract No. Collaborative
Research Center CRC-1044; Istituto Nazionale di Fisica Nucleare,
Italy; Ministry of Development of Turkey under Contract
No. DPT2006K-120470; National Natural Science Foundation of China
(NSFC) under Contract No. 11275189; U. S. Department of Energy under
Contracts Nos. DE-FG02-04ER41291, DE-FG02-05ER41374,
DE-FG02-94ER40823, DESC0010118; U.S. National Science Foundation;
University of Groningen (RuG) and the Helmholtzzentrum fuer
Schwerionenforschung GmbH (GSI), Darmstadt; WCU Program of National
Research Foundation of Korea under Contract No. R32-2008-000-10155-0.

\end{document}